
\input phyzzx
\nopubblock
\twelvepoint
\sequentialequations
\nonstopmode
\overfullrule=0pt
\tolerance=5000

\line{\hfill IASSNS-HEP-92/79}
\line{\hfill October 1992}
\titlepage
\title{QCD and Asymptotic Freedom: Perspectives and Prospects\foot{
Opening talk at the Aachen conference, ``20 Years of QCD'', June
1992}}
\author{Frank Wilczek\foot{Research supported in part by DOE
grant
DE-FG02-90ER40542}}
\vskip.2cm
\centerline{{\it School of Natural Sciences}}
\centerline{{\it Institute for Advanced Study}}
\centerline{{\it Olden Lane}}
\centerline{{\it Princeton, N.J. 08540}}
\endpage

\abstract{QCD is now a mature theory, and it is possible to
begin to view its
place in the conceptual universe
of physics with appropriate
perspective.  There is
a certain irony in the
achievements of QCD.  For the problems which
initially drove
its development -- specifically, the desire to understand
in detail the
force that holds atomic nuclei together, and later the desire to
calculate the spectrum of hadrons and their interactions
--  only limited
insight has been achieved.  However I shall argue that
QCD
is actually
{\it more\/} special and important a theory
than one had any right to anticipate.  In many ways, the
importance of the solution
transcends that of the original motivating problems.

After elaborating
these quasi-philosophical remarks,
I discuss two
current frontiers of physics that
illustrate the continuing
vitality of the ideas.

The recent wealth of beautiful precision
experiments measuring the parameters of the standard model has
made it possible to consider the unification of couplings in
unprecedented quantitative detail.  One central result emerging from
these developments is a tantalizing hint of virtual
supersymmetry.

The possibility of phase transitions in matter at temperatures
of order $\sim \nobreak 10^2~ {\rm Mev}$, governed by QCD dynamics, is
of interest from several points of view.  Besides
having a certain intrinsic
grandeur, this question:
does the nature of matter change qualitatively, as it is
radically heated? -- is important for cosmology, relevant to
planned high energy heavy ion collision experiments, and provides
a promising arena for numerical simulations of QCD.
Recent numerical work seems to be consistent with
expectations suggested by renormalization group analysis of the
potential universality classes of the QCD chiral phase transition;
specifically, that the transition is second order for two species of
massless quarks but first order otherwise.  There is an
interesting
possibility of long-range correlations in heavy ion collisions
due to the creation of large regions of
misaligned chiral condensate.

Finally at the end there is a brief discussion of
the relation between scaling violations and running of the
coupling.
Some
statements made later in the conference seemed to
indicate that the relationship
between these concepts is commonly
misunderstood, so I'm smuggling
this bit in even though it
wasn't part of the original talk.}

\endpage

\chapter{The Place of QCD and Asymptotic Freedom in the World-View
of Physics}

\section{`Practical' QCD}

The original goal of strong-interaction physics, dating from the
1930s, was of course simply to peel back one more layer in the structure
of matter -- specifically, to understand the forces holding atomic
nuclei together.

Ironically, QCD has not shed much light on the problems
that initially
motivated its development.
I have prepared a little table summarizing the
status of the applications of QCD:
\medskip
\tenpoint
\vbox{\tabskip=0pt \offinterlineskip
\def\tablerule{\noalign{\hrule}}
\halign to 5.5in{\strut#& \vrule#\tabskip=1em plus2em& #\hfil& \vrule#&
\hfil#\hfil& \vrule#& #\hfil&\vrule#\tabskip=0pt\cr\tablerule
&&\multispan5\hfil&\cr
&&\multispan5\hfil ``PRACTICAL'' QCD \hfil&\cr
&&\multispan5\hfil&\cr\tablerule\tablerule
&&\omit\hidewidth {\bf Experiments}\hidewidth&&
\omit\hidewidth {\bf Approximate Dates}\hidewidth&&
\omit\hidewidth {\bf Use of QCD}\hidewidth&\cr\tablerule\tablerule
&&&&&&&\cr
&&nuclear physics&&$1932\rightarrow$&&holy water&\cr
&&&&&&&\cr\tablerule
&&&&&&&\cr
&&resonance physics&&$1960\rightarrow$&&lattice
simulations&\cr
&&&&&&-still crude&\cr
&&&&&&&\cr\tablerule
&&&&&&&\cr
&&deeply inelastic&&$1968\rightarrow$&&1/2 solved&\cr
&&scatterings&&&&&\cr
&&&&&&&\cr\tablerule
&&&&&&&\cr
&&jet physics&&$1980\rightarrow$&&background!&\cr
&&(e.g. LEP)&&&&&\cr
&&&&&&&\cr\tablerule \noalign{\smallskip}
&\multispan5\hfil&\cr
}}
\twelvepoint
In many ways
it is the latest, technologically hardest and most expensive experiments
that we understand the best.
This of course is because of asymptotic freedom, which means that
simple behaviors can be expected at high energy.
In a funny way the most impressive triumph of
QCD is that it is now used by experimenters, designing
for future high energy accelerators, to estimate
their backgrounds.  A theory used to calculate backgrounds is
clearly a theory that people have faith in!

\REF\pilcher{For the application of QCD to jet physics,
see S. Bethke and J. Pilcher, {\it Tests of Perturbative
QCD at LEP\/} Heidelberg-Chicago preprint HD-PY 92/06 EFI 92-14,
to appear in {\it Annual Reviews of Nuclear and Particle Science\/}
{\bf 42} (1992).}

The techniques of perturbative QCD have been extended
to discuss not only
gross overall features such as the relative number
of 2-, 3-, 4-, and even 5-jet events, but
also more
refined aspects, including thrust and
especially angular distributions which clearly
exhibit the vector character
of the gluons [\pilcher ].
The ``Rutherford'' cross section for jet production
in pp collisions is probably the most
immediately striking result of this kind, but by no means
the only one, nor the one that can be most stringently compared
to theoretical calculations.

\REF\lam{I am here alluding to a large and rapidly developing
subject.  For an excellent review of the early work see
M. Mangano and S. Parke, {\it Physics Reports\/} {\bf 200}
301 (1991).  A recent reference for progress using the
string theory approach is Z. Bern and D. Kosower,
{\it Nuclear Physics\/} {\bf B379} 451 (1992).  Another very
interesting recent contribution is C. Lam, {\it
Navigating Around the Algebraic Jungle of QCD}, McGill
Preprint McGill/92-32 (1992).}

There has been remarkable progress over the last
few years in calculating complicated
processes involving many partons (quarks or gluons) at tree level.
This progress has come by combining several very ingenious tricks,
including brilliant use of
the special simplicity of helicity amplitudes in the spinor
formalism and systematic re-organization of the color factors [\lam ].
In this way calculations that at first seem frightening to contemplate,
such as the production of four gluons in a gluon collision,
become quite tractable.
Some of the ideas leading to these simplifications arise very
naturally in string theory.  Recently the techniques that
prove so powerful for trees
are beginning to be applied to loop diagrams also.
These developments are quite important for the future of QCD. As
a practical matter, multi-jet QCD processes form major backgrounds
to ``new physics'' searches,  and must be well understood quantitatively
if future accelerators (particularly the SSC) are to fulfill their
full potential.
It is also
quite hopeful that new qualitative insights
will emerge from
this re-organization of perturbation theory,
such as better understanding of exponentiation (Sudakov) and multiple
soft processes (non-abelian bremsstrahlung).

\bigskip

\REF\altarelli{Theoretical aspects of deep inelastic scattering
and other processes that may be compared quantitatively in QCD
are discussed in the talk by Altarelli in this volume.}

Working backwards in time up the list,
I think it
is fair to say that the problem of deeply inelastic scattering,
which was
decisive for the discovery of asymptotic freedom and the emergence of
modern QCD, is half solved [\altarelli ].

\REF\gw{D. Gross and F. Wilczek, {\it Physical Review\/}
{\bf D8} 3633 (1973); {\bf D9} 980 (1974);
H. Georgi and H. Politzer, {\it Physical Review\/} {\bf D9} 416 (1974).}

A new generation of high-statistics experiments (CCFR, BCDMS)
has allowed extremely precise
measurement of the evolution of structure
functions with increasing $Q^2$.
The pattern of the scaling violations is
exactly what was predicted in 1973 [\gw ].  They
provide  quantitative tests of  predictions from
a strongly
interacting field theory -- indeed, the field theory
of the strong interaction --
at a level which would have been unthinkable before that
time, and was barely conceivable then.
(I remember vividly the incredulity
and
near-ridicule the predictions initially met. A famous
experimentalist challenged me:
``You expect us to measure logarithms?
Not in
your lifetime, young man.''  So it is gratifying now, nineteen
years later, to come to this conference and find that the third
decimal place in $\alpha_s(M_W)$ is being debated.)
We eagerly await results from HERA, which
will vastly extend the scope of these measurements.

\REF\brodsky{See the talk by Brodsky in this volume.}

\REF\sachradja{See the talk by Sachrajda in this volume.}

By way of contrast
the theory of the starting structure functions remains
at a comparatively
primitive stage.  Promising ideas from light cone field theory are
maturing [\brodsky ], and
lattice workers are getting closer to
the goal of
simulating the continuum theory with
light dynamical quarks [\sachradja ], so there are grounds
for optimism.

Similar remarks apply to the study of
form factors and the rich set of phenomena
associated with the Drell-Yan process
and heavy quark production,
which are other half (or maybe slightly less than half)
solved problems.

In many ways deep inelastic
scattering
provides a cleaner
quantitative test of QCD than jet physics.  This is
because in deep inelastic scattering -- in and a few other
totally inclusive processes (see
below) -- the predictions
can be derived within a theoretical framework where one has a fairly
precise idea of what the errors are.   In any but totally inclusive
processes one has to face up to the problem of matching practical
calculations involving quarks and gluons to the hadrons that
experimentalists in the end observe.  The art of ``hadronization''
algorithms
has become highly developed, and extremely impressive fits to
the data over a wide range of experiments at different energies have
been achieved.  Still, it is difficult to estimate the errors.

\REF\braaten{E. Braaten, S. Narison, A. Pich {\it Nuclear Physics\/}
{\bf B373}  581 (1992).}

The total hadronic
cross-section in $e^+~e^-$ annihilation and the hadronic
branching ratio in $\tau$ decay are totally inclusive in a
stronger and more obvious way even than deep inelastic scattering,
since for them there is no target proton or nucleus.
In a bold yet careful analysis Braaten, Narison, and Pich
have made a plausible case that $\tau$ decay affords an especially
discriminating determination of $\alpha_s$.  At first hearing
this claim seems quite startling, but on reflection its logic
is compelling.  There are two basic reasons why $\tau$ decay is
favorable.  The first is very general: while it starts out
large at low energy
the effective coupling becomes small, and runs slowly, at high energy.
Thus at high energy
its effects are small, and to see running of the coupling
directly at high energy one must discern small changes in small effects.
(See however the final section of this talk.)
As Altarelli emphasizes, this is neither a waffle nor a barrier
to testing asymptotic freedom, but rather a very strong positive
statement: from comparatively crude determinations of $\alpha_s$ at
low energy one gets precise predictions at high energy, with no
free parameters at all.  On the other hand it puts a premium on clean
low energy determinations, such as from $\tau$ decay.
The second reason
is more special.  Although it is slightly technical, I
will mention it
here to make a point.
It happens that the particular weighted
integral of current correlation functions which governs $\tau$ decay
is quite clever:  it kills -- {\it via\/} Cauchy's integral
theorem, since the term of the requisite power vanishes --
the contribution of the lowest dimension
operators (dimension 4) that correct the lowest order result
from the identity operator, and are
not very precisely known.  The point I want to make is that we
should aspire to being
as clever as three-body phase space in other problems
of this kind.  It ought to be possible to take appropriate integrals
over the $e^+~e^-$ data, or over $Q^2$ in deep inelastic scattering, to
suppress the poorly known contribution of low-dimension or low-twist
operators, and thereby to obtain comparably
precise determinations
of $\alpha_s$ at low energies from these processes.

\bigskip

Moving further up the table
we come to resonance physics.  A host of techniques
have been proposed to meet the challenge of connecting the microscopic
theory of QCD, which of course is formulated
in terms of degrees of freedom that do not
appear in the physical spectrum, to the rich material of the Rosenfeld
table.  These techniques include most impressively the
strong coupling expansion, which played a crucial historical role
in providing important
evidence for the idea of confinement and supplies a
surprisingly good (and I think insufficiently appreciated)
semi-quantitative version of the naive quark model; and the
large $N$ expansion, which rationalizes several striking qualitative
features of QCD dynamics.
There are also of course a number of less ambitious
semi-phenomenological
approaches including the M.I.T. bag model, QCD (ITEP) sum rules,
Regge theory, chiral perturbation theory, and others, that are extremely
useful but do not purport to be complete or systematically improvable.
For serious quantitative comparison
of theory and experiment none of these approaches
is adequate.  At present it appears that
there is no alternative to the brute-force approach
of lattice gauge theory -- that is, to
doing the functional integrals numerically.
Progress in this field has been steady.
A convenient if imprecise yardstick of progress in the field is
the deviation of the calculated value of $m_p / m_\rho$ from the
naive quark  model value 3/2.  One will obtain a value close to
3/2 if the effective quark mass is large.  This can happen either
because the quarks in the simulation have a large intrinsic mass
or because they are localized on a small lattice, thus inducing
{\it via\/} the uncertainty principle
large kinetic energy which dominates their interaction energy.  Existing
simulations suffer from one or both of these problems: it is difficult
to take the quark mass to zero, because as the range of the quark
propagator (correlation length) increases the calculation slows down; and
it is difficult to take a large lattice for the same reason.  For these
reasons, only very recently have the lattice calculations given an
answer for $m_p / m_\rho$ distinguishable from 3/2, and they
are still rather far from the physical value $\sim 1.2$.  The
difficulties here are not matters of principle, and they will be
resolved as a matter of course as available
computing power increases.   This will be a slow incremental process,
however, in the absence of new ideas.
We may have to wait another decade before
10\% accuracy is achieved in this or similar applications.

\REF\elk{A. El-Khandra {\it et al.} Fermilab-pub 91/354-T
(1991).}

Determination of heavy quark potentials and
interactions appears much less demanding.  There are already
significant attempts to compare lattice results for heavy quark
physics to observation, including a respectable determination of
$\alpha_s$ [\elk ].
I expect such calculations will soon afford the most
accurate determinations of $\alpha_s$.

\bigskip

Finally, as far as I am aware the influence of
microscopic QCD within its most important niche in the natural
world, that is the physics of atomic nuclei,
has been quite marginal. (I would love to be corrected on this point.)
It can be used to sprinkle holy water on the Skyrme model, which can
be a useful semi-phenomenological tool.  Also there are some beautiful
effects in perturbative QCD that show up for nuclear targets,
such as the disappearance of shadowing at large $Q^2$ -- but these are
clearly marginal to nuclear physics proper.  The basic problem
in relating nuclear physics to QCD is similar to the the basic
problem in relating chemistry to atomic QED: there is a big mismatch of
energy scales.  The fundamental scale of QCD is at least
$100 ~{\rm Mev}$, whereas nuclear physics in concerned with energies
one-tenth or one-hundredth of this.
In fact
probably the most important contribution that QCD makes to
nuclear physics is
to tell us {\it not\/} to seek new fundamental laws in that domain.
QCD focuses attention in nuclear physics
where it belongs, toward the construction of useful phenomenological
models and toward the challenge of relating qualitative phenomena,
especially those which do not rely on delicate
energy differences, to a known
microscopic theory.  Can one, for example, derive the hard-core
interaction or the saturation of nuclear forces from QCD in a convincing
fashion?

\bigskip

An important philosophical point that emerges clearly from this
brief
survey of `practical' QCD is that {\it in the end,
the theory itself informs us which phenomena are simple and
fundamental, and
which are intrinsically complex and secondary.}
This is typical of many problems in physics, and for that matter
other branches of science.
Bohr recalled that at the time he proposed
his famous model
atomic spectra were viewed ``like the pattern of butterflies'
wings,'' as beautiful but secondary and hopelessly complicated
manifestations.  Of course Bohr's model changed this perception --
for hydrogen.  It took many years and
several new insights to partially decode
the more intricate spectra of larger atoms and molecules, and in fact
apart from a few salient regularities they {\it are\/} rather analogous
to the pattern of butterflies' wings.
In QCD deeply inelastic scattering, a few
other hard processes, and some aspects of
heavy quark
physics yield the observables
most closely related to simple and fundamental
parts of the theory.
For the foreseeable future they will form the arena
wherein the theory is most easily tested and put to use quantitatively.
The grand old problems of nuclear and resonance physics are still with
us, but are seen in a different light --  not as potential sources
of conceptual crises, but as
challenges to ingenuity and deduction.

So much for the `practical' applications of QCD.
They offer more than sufficient proof of the correctness of the theory,
but
limited insight into the original problems which motivated it.
In my opinion, despite its
practical limitations QCD is a much more special and important
theory than one might have anticipated from its origins.  This is
because it either
directly precipitated or helped to catalyze three conceptual
revolutions.

\section{First revolution: quantum field theory is incarnated}

The development of QCD and asymptotic freedom
changed the way
people regard quantum field theory.  It made it clear that
{\it one must take quantum field theory, including its ultraviolet
problems and its non-perturbative aspects, deadly seriously.}
This attitude is now so deeply
ingrained that it may be difficult for
young people who missed experiencing it,
or older people with fading
memories, fully to imagine the intellectual atmosphere in the
1960s and early 1970s,
when it was an extremely unfashionable one.  It is quite
instructive to look into the literature of those times.
Many if not most
theoretical papers dealing with the strong interactions
contained an obligatory ritual mantra wherein the S-matrix or
bootstrap was invoked, before getting down to their actual point
(often rather tenuously connected to those theological principles).
Use of strict
quantum field theory was considered to be naive, in rather poor taste,
an occasion for
apology.
One hears an echo of these attitudes
even in the conclusion of Gell-Mann's
famous 1972 summary talk at the NAL conference:

``Let us end by emphasizing our main point, that it may well be
possible to construct an explicit theory of hadrons, based on quarks
and some kind of glue, treated as fictitious, but with enough
physical properties abstracted and applied to real hadrons to
constitute a complete theory.  Since the entities we start with are
fictitious, there is no need for any conflict with the bootstrap or
conventional dual parton point of view.''

What were the reasons for this suspicion
of quantum field theory,
which in retrospect appears strange?  Part of the
reason was historical.  The late 1940s and early 50s saw
what appeared on the face of it to be a great triumph for
quantum field theory, the triumph
of renormalization theory in QED.  However
the procedures developed at that time
for solving, or even making sense of,
the
equations of QED
were intrinsically tied to a perturbative
expansion in powers of the coupling constant.  For
QED this coupling is indeed small, but in the
then-current candidate
quantum field theory of the strong interactions, Yukawa's
pi meson theory, it was clear that the coupling would have
to be
large for the theory
to have any chance of agreeing with experiment.  Thus although
this theory was not known to be wrong, it was certainly
useless in practice.  Attempts to solve the theory without
resorting to perturbation theory did not succeed, both for
practical reasons and for a fundamental one
that we will discuss momentarily.

As the rich phenomenology of resonance physics was discovered,
theorists for the most part
made progress toward digesting it not by the top-down
approach of deriving mathematical consequences from a powerful
fundamental theory, but rather by more modest methods based on
symmetry and high-class kinematics.  (In
the category of high-class kinematics I
include dispersion relations, derived from causality, and
S-matrix
model-building guided by pole-dominance or
narrow-resonance approximations together with the
constraint of unitarity.)

Thus quantum field theory gradually lost much of its lustre.
The successes of field theory in QED were rationalized as
due to
a lucky accident.  One could to a certain extent recover
these successes from the less committal point of view
fashionable
in strong interaction physics, along the following lines:
the weak coupling expansion of
quantum field theory is essentially a systematic way of unitarizing
the single pole amplitude for photon exchange, supplemented with
the assumption that the relevant dispersion relations need no
subtraction.
This philosophy appeared especially sensible given that renormalization
theory failed even for the other available weak-coupling theories
of the weak interactions and of gravitation, while the modest
semi-kinematic approach worked perfectly well in these domains,
and was extremely fruitful
in untangling the weak interactions of hadrons.

\REF\landau{L. Landau, in {\it Niels Bohr and the
Development of Physics}, ed. W. Pauli.  (McGraw-Hill,
New York 1955).}

But the difficulties in accepting quantum field
theory at face value were not only matters of history and
sociology.  The only powerful method for
extracting consequences
from non-trivial interacting quantum field theories
was perturbation theory in the coupling.  This perturbation
theory, implemented in a straightforward way, gave infinite
results order by order due to the exchange of highly virtual
quanta.  Tomonoga, Schwinger, and Feynman, building on
qualitative insights of Kramers and Bethe, were able to
make sense of the perturbation theory term by term,
using a tricky limiting procedure
that in modern terms amounts
to expressing the perturbation theory in terms of the
effective coupling at a small momentum typical of the
physical situation considered.
The convergence of the perturbation theory, upon
which the renormalization procedure hinged,
was very doubtful (in fact it fails to converge
for almost any non-trivial
theory, though in favorable cases
can be rescued by Borel resummation.)
What now appears to be the most profound point was made by
Landau [\landau ].  In modern language, his point was that in a
non-asymptotically free theory the coupling instead of
decreasing logarithmically at small distances would increase,
and inevitably become large.  Thus the procedure of expanding
in a small low-energy effective coupling only hid but did
not remove the inevitable appearance of strong couplings among
the virtual quanta, which invalidate the perturbation series.
Indeed the fundamental bare coupling, which
to satisfy the requirement of locality in a theory of particles
must be fixed at infinitely small separations,
formally diverges to infinity.
If one defines the theory by a regularization or
cut-off procedure, which roughly speaking corresponds to
specifying the coupling at a small but finite distance and
letting this distance become
smaller and smaller while adjusting the
coupling accordingly, then to obtain finite results
at finite
distances the bare coupling must be taken to zero.
But doing that, of course,
leads to a trivial, non-interacting theory.  Landau's argument
that non-asymptotically free theories cannot exist
is not rigorous, because the logarithmic running of the
coupling on which it is based can only be derived at weak coupling.
(It is a fully
convincing argument that such theories
cannot be constructed {\it perturbatively}.)
Yet later work in ``destructive field theory'' has largely
vindicated Landau's intuition, and showed that many theories,
almost certainly including QED and Yukawa's pion theory, in fact
do not exist (or are trivial) despite the fact that their
perturbative expansions are non-trivial term by term.

Developments in the late 1960s and early 70s put these issues
in a new light.  The successful use of non-abelian
gauge theories to construct models for
the electro-weak interactions, and 'tHooft's proof of
their renormalizability, provided a wider perspective in which
to view the earlier success of QED.  They made it seem
less plausible
that the successful use of quantum field theory in QED was a lucky
fluke.  They also raised the possibility that the unification of
electrodynamics with other interactions would cure its most severe
fundamental problem, the Landau problem just described.

\REF\partons{R. Feynman {\it Physical Review Letters\/}
{\bf 23} 1415 (1969);
J. Bjorken and E. Paschos {\it Physical Review\/} {\bf 185} 1975 (1969).}

On the other hand, the success of the quark-parton model in
describing the results of the SLAC
deep inelastic electroproduction
experiments created a rather paradoxical situation for the
theory of the strong interaction.  The quark-parton model was
based, essentially, on an intuitive but not wholly consistent
use of {\it non-interacting\/}
field theory for the supposed constituents of
strongly interacting hadronic matter.  Landau's argument
was meant to be a {\it reductio ad absurdum\/} -- showing that
the only consistent quantum field theories must be non-interacting
at short distances, and therefore trivial.  It seemed as if Nature
accepted Landau's argument, but failed to draw the obvious, absurd
conclusion!  This craziness,
together with the vulgar problem that the
quarks were never observed as individual particles, helped foster
that skepticism
both of quantum field theory and of the real existence
of quarks, which Gell-Mann expressed so eloquently.

Such was the intellectual climate in which Gross and I
began our
discussions in 1972.  He had been grappling
with the SLAC results for several years, and felt very keenly
an ever heightening tension
between the phenomenological success of simple
field theory ideas in describing those results and the rotten
foundation of those ideas.  He was hoping to show a definite
contradiction, so that a clean break with the past could be
justified (analogous to
Bohr's clean break with classical physics in his
atom model).   I was a graduate student mainly interested in
the weak interactions and enormously impressed by
the gauge theories of electroweak interactions, which were
developing rapidly and displayed
a mixture
of beauty and awkwardness that
was fascinating -- especially to a graduate student looking
for problems.  I
was eager to understand how
these theories behaved at high energies,
and in particular whether they could manage to avoid the
Landau problem.

Exaggerating only slightly, I think it would be fair to say
that David was trying to demolish quantum field theory as
a possible description of the
strong interaction, while I was hoping to vindicate it for
the weak interaction.  From both these points of view, the results
of our work were negative.

We found, as you know, a crucial reversal of sign in the
change of effective coupling with momentum scale occurs
in non-abelian gauge theories.
This
indicates that the coupling
shrinks instead of growing at short distances: asymptotic
freedom.  Such
behavior is rather counter-intuitive, and had not
been encountered in other quantum field theories.
It had not been anticipated, by Landau or anyone else.
It allowed one both to
circumvent Landau's argument
and to understand qualitatively --
and, soon, quantitatively -- the SLAC results.
We immediately
recognized that the degrees of freedom that naturally
occurred in asymptotically free theories were precisely those
one needed to construct a model of (colored) quarks interacting
with (colored) gauge gluons, though this
had {\it not\/} guided
our search.
It thus became evident that quantum field theory
might
supply a consistent theory of the strong interaction, after all
-- and soon we were convinced it did.
On the other hand we found, by painful examination of many models,
that the scalar Higgs fields needed in gauge theories of the
weak interactions made it very difficult to maintain asymptotic
freedom.  We did not find any examples where one had enough
Higgs fields to break the
symmetry completely, or down to an abelian group,
without losing asymptotic freedom and reinstating the Landau problem.
Thus field theory as applied to electroweak physics remained
fundamentally problematic.

\REF\mueller{See the talk by Mueller in this volume.}

However, I can't say we were terribly
disappointed.  We had found, for the first
time, a class of interacting,
relativistic quantum field theories in four space-time
dimensions
that had a reasonable chance of actually existing.
(In this regard, after the previous discussion
it is perhaps worth mentioning that
an infrared version of Landau's problem, the inevitable
occurrence of a strong coupling among very soft
virtual quanta, does occur in QCD.  However one expects that
the development of non-trivial vacuum structure removes
any true divergence; how this may happen
can be very crudely conceived by imagining that
gluons obtain an effective mass.
Infrared renormalons [\mueller ], which
complicate the interpretation of perturbation theory but
presumably do not endanger the theory as a whole, are the
surviving tangible
residue of Landau's argument.)

Over the past nineteen years an ever widening
network of successful quantitative tests and applications of
perturbative QCD has grown, as this conference amply attests.
The old ``problem'' of infinities in perturbation theory, and
the effects of highly virtual particles which give rise to them,
lie close to
the very root of all these successes.  Whereas in QED and
electroweak theory higher order effects of quantum
field theory generally
provide small corrections,
in QCD they are much larger quantitatively.
Two- and even three-loop calculations are needed to
address the data adequately.  And of course, the logarithmic
infinities due to highly virtual quanta of large invariant mass
are directly responsible
for the running of
the coupling, which is the conceptual foundation of all QCD
phenomenology.
Thus to the extent that the predictions of QCD perturbation
theory are verified, the detailed structure
of quantum field theory and
its renormalization program stand dramatically vindicated.

\bigskip

The phenomenological success of QCD and asymptotic freedom
in describing a wide variety of hard processes using souped-up
methods of perturbation theory does not, of course,
remove the challenge of understanding non-perturbative aspects of
the theory.  Perhaps nothing exhibits this challenge so clearly
as the fundamental formula of dimensional transmutation.  QCD
at the classical level contains only a
dimensionless coupling and is scale invariant \foot{Strictly
speaking this holds only for massless quarks, but the essence
of the following argument is valid generally.}
On the other hand physical hadrons have definite,
non-zero masses.
How does a
parameter with dimensions of mass emerge from a fundamentally
massless theory? It happens because the running of the coupling,
which is an inevitable result of quantizing the theory,
implicitly defines a mass scale:
$$
\Lambda ~=~ \lim_{Q^2 \rightarrow \infty} Q e^{-c_1/\bar g^2(Q^2 )}
   [g^2 (Q^2 ) ]^{c_2}~.
\eqn\runcoup
$$
Here $c_1$ and $c_2$ are definite numbers that can be read off
from the first two terms in the
renormalization group $\beta$ function.  The main
point is that the limit
on the right-hand side exists and defines a finite mass.  All
other masses in the theory, including the masses of particles in
the spectrum, can be expressed as {\it pure numbers\/} times this
one.  Once the boundary condition for the running coupling
is determined the theory is completely fixed, there is no other
remaining parameter nor any
independent scale.  For our present discussion the
most significant
point is that all hadronic masses therefore will be, like
$\Lambda$, non-perturbative in $g^2$.  The challenge could
not be clearer.

\REF\creutz{An excellent review of the principles of
lattice gauge theory, by the theorist who first carried
through the argument just mentioned, is M. Creutz,
{\it Quarks, Gluons,
and Lattices\/} (Cambridge, 1983).}

\REF\largeN{G. 'tHooft, {\it Nuclear Physics\/} {\bf B72}
461 (1974); E. Witten {\it Nuclear Physics\/} {\bf B160}
57 (1979).}

The challenge of understanding non-perturbative effects in QCD
has led to several remarkable developments.
Perhaps the most important single result is that there is
now a convincing
case that the microscopic theory of QCD actually does give rise
to the confinement of
quarks and gluons inside hadrons.
One demonstrates this by showing,
using computer simulation,
that there is no qualitative change (phase transition) between
strong-coupling expansion of the discretized lattice
theory, in which
confinement is manifest but Lorentz invariance is violated --
and the
continuum limit, in which
Lorentz invariance is manifest but confinement
is not [\creutz ].  The application of
Monte Carlo methods, semiclassical approximations,
and large $N$ expansions [\largeN ] for quantum field theories
have been highly developed,
with QCD as one of the important original
motivations but now ramifying into many other areas.

\REF\tHooft{G. 'tHooft, {\it Physical Review Letters\/}
{\bf 37} 8 (1976); C. Callan, R. Dashen, D. Gross {\it
Physics Letters\/} {\bf 63B} 334 (1976); R. Jackiw and
C. Rebbi {\it Physical Review Letters\/} {\bf 37} 172 (1976).}

\REF\pq{R. Peccei and H. Quinn, {\it Physical Review Letters\/}
{\bf 38} 1440 (1977).}

\REF\axion{S. Weinberg, {\it Physical Review Letters\/}
{\bf 40} 223 (1978); F. Wilczek, {\it Physical Review Letters\/}
{\bf 40} 279 (1978).}

A particularly
striking discovery
is the possibility of non-perturbative P and T violation in
QCD: the famous $\theta$ term [\tHooft ].  What seems to be the
most satisfactory approach to
understanding why this potential source of P and T violation is
in fact highly suppressed (as shown by the smallness of the
neutron's electric dipole moment) was suggested by Peccei and
Quinn [\pq ].  It involves the existence of a new light boson, the
{\it axion}, with remarkable properties [\axion ].
If axions do exist, they
may be very important for cosmology, plausibly even supplying the
astronomer's
``missing mass'', which is about 90\% of the Universe by weight.

\bigskip

The development of QCD had a curious effect on string theory.
Its immediate impact was certainly to
kill much of the interest in string theory,
which of course was originally developed as a model
of the strong interaction.  By providing a correct microscopic theory
of the strong interaction based on quite different principles --
and incorporating in a central place
point-like interactions at short
distances that are quite difficult to reproduce in a theory
containing only extended objects -- QCD removed from string theory
its initial source of motivation.  For the longer term however the
story is more complicated, and its conclusion is not yet clear.
By emphasizing that the short-distance properties of quantum field
theory must be taken deadly seriously, and that the ``problems''
encountered in perturbation theory are not
mere mathematical artifacts but rather signify deep properties of
the full theory, the development of QCD made the corresponding --
apparently intractable -- problems encountered in the perturbative
expansion of Einstein gravity seem that much more weighty.  Thus
the discovery that string theories can incorporate Einstein gravity
while avoiding its bad short-distance behavior is taken as
a powerful argument in favor of these theories.

\section{Second revolution: unification becomes a science}

To achieve a unified description of apparently vastly different aspects
of Nature is certainly a major esthetic goal of the physicist's quest.
In the past it has also been a fruitful source of essentially new
insight: Maxwell's fusion of electricity and magnetism
transformed our understanding of optics, and vastly generalized it;
Einstein's fusion of special relativity with gravitation transformed
our understanding of space-time and cosmology.

\REF\gg{H. Georgi and S. Glashow, {\it Physical Review Letters\/}
{\bf 32} 438 (1974).}

\REF\gqw{H. Georgi, H. Quinn, S. Weinberg {\it Physical Review
Letters\/} {\bf 33} 451 (1974).}

The development of QCD and asymptotic freedom has enabled us
to add a major new chapter to the story of unification.  There
are two aspects to its contribution.  First,
{\it the mathematical resemblance of QCD to the gauge theories
of weak and electromagnetic interactions immediately suggests the
possibility of a larger gauge theory encompassing them all}.
Georgi and Glashow [\gg ]
constructed a compelling model of this kind
almost before the ink was dry on asymptotic freedom.
Second, the running of couplings removes the major
obvious -- superficial -- difficulty in the way of implementing such
an extended gauge symmetry, that is the disparity of coupling strengths
as observed at accessible energies.  Georgi, Quinn, and Weinberg
[\gqw ]
showed how to use the renormalization group as a quantitative tool in
investigating unification.
{\it The running of the couplings makes it possible
to study ambitious unification schemes quantitatively, and
compare them to observations.}
The remarkable success of this line
of thought, and its recent exciting development,
deserve a section in
themselves (below).

\section{Third revolution: the early universe opens to view}

\REF\wein{S. Weinberg, {\it Gravitation and Cosmology\/}
(Wiley, New York 1972).}

The position of very early universe cosmology just prior
to the discovery of asymptotic freedom is well conveyed in
Weinberg's classic text [\wein ] (1972):

``However, if we look back a little further, into the first 0.0001 sec
of cosmic history when the temperature was above $10^{12}~$K, we
encounter theoretical problems of a difficulty beyond the range of
modern statistical mechanics.  At such temperatures, there will be
present in thermal equilibrium copious numbers of strongly interacting
particles -- mesons, baryons, and antibaryons -- with a mean
interparticle distance less than a typical Compton wavelength.  These
particles will be in a state of continual mutual interaction, and cannot
reasonably be expected to obey any simple equation of state.''

This pessimistic picture
changed overnight when the discovery of asymptotic freedom.
Instead of being mysterious and intractable, matter
at extreme temperatures and densities
becomes weakly interacting and its
equation of state simply calculable.  This
development, together with the ideas of unification just mentioned,
opened up a vast new field of investigation.
{\it It becomes possible to
make reasonable guesses for the behavior of matter under much more
extreme conditions\/}
than the mere $10^{12}~$K mentioned by Weinberg, and
to calculate the consequences of various unification scenarios for
cosmology with
some confidence.

\REF\kt{For a review of very early universe cosmology
see E. Kolb and M. Turner, {\it The Early Universe\/}
(Addison Wesley, Redwood City 1990).}

What has emerged from this opening [\kt ]?
There is now at least one plausible
scenario, based on baryon-number violating interactions
at the grand unified scale,
for how the asymmetry between matter and antimatter could
have developed from a symmetric starting condition.  Much
recent work has been devoted to the possibility of developing
such an asymmetry at the weak scale, though
the viability of this idea is presently unclear.
Various unification models can be constrained cosmologically
(for example if they create stable domain walls, contain too many
massless neutrinos, or contain
stable particles that would be produced in the
big bang in sufficient abundance that they could not have
escaped notice, ...).  Most exciting, plausible
candidates for the ``dark matter'' or ``missing mass'' have emerged.
These are particles that -- given their existence --
one can calculate would have been produced
in the big bang in sufficient abundance to redress the mismatch
between the density of ordinary matter observed directly or inferred
from nucleosynthesis and the amount necessary to account for the
gravitational dynamics of galaxies and clusters.  It is also necessary,
of course, that the postulated particles would have escaped observation
to date.  It is remarkable that two specific
kinds of particles: axions as
mentioned above, and LSPs (lightest supersymmetric particles) which
arise naturally in supersymmetric unification as discussed below,
seem to fit the
bill.  Heroic experiments are proposed to search for these particles,
whose (very different) properties are reasonably definitely predicted.
It would be difficult to overstate the importance of
a positive detection.
Finally, an impressive circle of ideas
around inflation has developed.

It would be ridiculous to claim that QCD
and asymptotic freedom are directly
responsible for all these developments,
which
require many new independent ideas.  Besides,
the ultimate value of
these specific very speculative ideas
can't yet be reliably assessed.
I believe, however, we may already conclude
that the once seemingly impenetrable
veil of ignorance described by Weinberg,
which appeared to separate us
from sensible scientific
contemplation of the earliest moments
of the big bang, will
never again seem
so formidable.  Truly ``we live in the age of the trembling of
the veil.''

\chapter{Unification of Couplings}

\REF\dim{S. Dimopoulos, S. Raby and F. Wilczek,
{\it Physics Today\/} {\bf 44}, October, p. 25  (1991)}

(Since I have recently written on this subject
at roughly the level of the talk, I shall be very
brief here, directing you to [\dim ]
for more details and a full set of
references.)

The logic that enables one to
connect unification ideas
quantitatively with low-energy observations
is as follows.  One observes three {\it a priori\/}
independent couplings, corresponding
to the three gauge groups $SU(3)\times SU(2)\times U(1)$ of
the standard model, at low energies.  In a unified theory
these couplings are in reality not independent, but derive
from a single coupling.  The difference between their
observed values at low
energies must be ascribed to the different evolution of
the respective running couplings down from the energy
scale of unification.  The running of these couplings
is basically determined by the particle content of the
theory, given two inputs: the energy at which the large
gauge symmetry broke (often called the GUT scale), and the
value of the coupling at that scale.  Since therefore three
observed parameters arise from two input parameters, they
are {\it overconstrained}.  Given a specific unified
model, the
constraint may or may not be met.
If it is not met, we must discard the model.  If it is
met, then
that fact is
a highly non-trivial success for the model and
for the
assumptions that go into the calculation.

In connection with unification it is profoundly
important that the couplings run slowly;
that is, logarithmically
with energy scale.
Since there is a big discrepancy between the
effective strong
and weak couplings at presently observed energies,
there are factors
of the type $e^{\kappa\over \alpha}$ relating current
accessible scales to the unification scale.  In typical
models the GUT scale turns out to be of order
$10^{15}-10^{17}$ GeV.  This mass sets the scale for
exotic processes that occur through exchange of gauge
bosons which are in the unified group but not in
$SU(3)\times SU(2)\times U(1)$, including proton decay.
Also its large value is definitely smaller than, but
not incommensurate with, the
Planck energy $M_{\rm Pl.}\approx 10^{19}$ GeV.
where the gravitational
interaction becomes strong.  This closeness provides
hints at
an organic connection between gravitation
and traditional particle physics.  On the other
hand the fact that
the GUT scale is significantly smaller than
$M_{\rm Pl.}$ makes it plausible that we can calculate
the running of the couplings all the
way to unification without encountering significant
corrections from quantum gravity.

Until recently the minimal
unified model, based on the unifying gauge group
$SU(5)$, gave an adequate fit to the data.  That is,
the observed couplings satisfied, within their quoted
uncertainties, the constraint derived in the manner
described
above for this model.
This represents a truly extraordinary triumph for
quantum field theory, extrapolated far far beyond
the domain of phenomena it was designed to describe.
It also might seem at first sight to be rather depressing,
since it suggests a vast ``desert'' between present
energies and the GUT scale.  To be more precise: if we
do not believe the success of this calculation to be an
accident, we must not only take unification seriously, but
also make sure that unification schemes more elaborate
than
the simplest possible one
manage to give something close to the same answer.
Those intent on populating the desert
-- or (techni-\nobreak)colorizing it -- should be required to
submit an appropriate environmental impact statement!

Recent, beautifully accurate
measurements of standard model
parameters from LEP and elsewhere have made it clear
that actually minimal $SU(5)$ doesn't quite work.  The
observed couplings are close to satisfying its constraint,
but the discrepancy is now well outside the error bars.

There are various possibilities for addressing the
discrepancy, among which one seems especially noteworthy.
The noteworthy possibility is that the quantitative study
of unification of couplings has uncovered evidence for
virtual supersymmetry.

There is a standard litany of the virtues of supersymmetry,
probably familiar to you all: it enables a new level of
unification, between particles of different spin; it
ameliorates the gauge hierarchy problem (see below); it
is necessary to eliminate tachyons in superstring theory.
But in any list of the
virtues of supersymmetry, one entry is conspicuously
absent: experimentally verified consequences.

How does supersymmetry affect the running of the couplings?
It might seem at first glance that its effect is bound to
be catastrophic, since it roughly doubles the particle spectrum.
It might seem that all these new virtual particles would
inevitably induce a drastic change from the
nearly successful results
for minimal $SU(5)$.
However, it is an important fact that adding {\it complete\/}
$SU(5)$ {\it multiplets\/} to the theory affects the calculation
of the constraint among observed couplings arising
from unification of the couplings only very little.
This is because, roughly speaking, virtual particles forming
such complete
multiplets affects all three couplings in the same way.
They change the value of the
unified coupling
at the GUT scale, and can slightly modify the size of that
scale, but to a good approximation they leave the constraint
among observed couplings unchanged.

Since supersymmetry is basically a space-time as opposed to
an internal symmetry, in extending the minimal unification scheme
to incorporate supersymmetry one simply doubles all the complete
multiplets that were in the original model, by adding their
supersymmetric partners.  The gluinos do not occur in
complete multiplets, but their contribution has the
same structure as that of the ordinary gluons,
and therefore they do not alter the
group-theoretic
structure of the calculation. (They do significantly
alter the
predicted GUT scale and coupling.)  The Higgs multiplets are a
different story, however.  The Higgs particle in the standard
model is {\it not\/} part of a complete $SU(5)$ multiplet at
low energies; its color triplet partner is capable of mediating
proton decay, and must be extremely heavy.  There
is no convincing theoretical
explanation of why it should be -- this is one aspect
of the gauge hierarchy problem.  In passing to the supersymmetric
version of the minimal unified model one must add the fermion
partners of this standard model doublet.  In fact, for slightly
subtle reasons, one actually must add
two such Higgs complexes, for it is
impossible to maintain supersymmetry with only a single Higgs field
giving masses to both up and down quarks.

Thus in the minimal
supersymmetric model one must add quite a few fields that do not
form complete $SU(5)$ multiplets and do affect the constraint among
low-energy couplings.  Remarkably, when this is done the
modified prediction agrees with the accurate modern experiments.

If we take this agreement at face value, as an indication for
the effect of virtual supersymmetry, it augurs a bright future
for experimental high energy physics.  If supersymmetry is to
fulfill its natural role in ameliorating the gauge hierarchy problem,
it cannot be too badly broken.  Specifically, if the the cancellation
between virtual particles and their supersymmetric partners is
not to generate corrections to the
Higgs mass that are formally larger
than that mass itself, the generality of
superpartners cannot be much heavier than
$M_W /\alpha~\approx 10~{\rm TeV}$.  Some are expected
to be considerably
lighter.  Thus they fall
within the range of foreseeable accelerators.
Also, although supersymmetric unification raises the GUT scale and
thus decreases the rate for proton decay by exotic
gauge boson exchange,
it does not do so by an enormous factor.  The predicted
range of rates,
although safe from existing bounds, does not seem hopelessly out
of reach.

\chapter{The Phase Transitions of QCD}

\REF\phase{An interesting discussion of possible
phase transitions in QCD is given in E. Shuryak,
{\it The QCD Vacuum, Hadrons and Superdense Matter\/}
(World Scientific, Singapore 1988).  For recent progress
one may consult {\it Lattice '90}, {\it Nucl. Phys.}
{\bf B (Proc. Suppl.) 20} (1991); {\it QCD '90}
{\it Nucl. Phys.} {\bf B (Proc. Suppl.) 23} (1991);
{\it Quark Matter '90} {\it Nucl. Phys.} {\bf A525}
(1991) and their successors.}

\REF\sy{B. Svetitzky and L. Yaffe, {\it Physical
Review} {\bf D26} 963 (1982).}

\REF\kw{The discussion in this section
from here on roughly follows a much fuller
one in
K. Rajagopal and F. Wilczek, {\it Static and Dynamic
Critical Phenomena at a Second Order QCD Phase Transition\/},
Princeton-IAS preprint PUPT-1347, IASSNS-HEP-92/60; and
further work to appear.}

The possibility of phase transitions [\phase ]
in QCD is
fascinating in itself.  We are asking, what
happens to matter if we heat it to very extreme
temperatures?  In addition it is of interest for
cosmology, since the requisite temperatures would
have been achieved in the early moments of the big
bang; for numerical experiments, since
(homogeneous) thermodynamic
quantities are among the easiest to measure and
interpret in lattice gauge theory simulations; and
for heavy ion collisions.

Actually the questions posed by each of these applications
are rather different.  In the numerical experiments one can
easily imagine varying the number of quarks or their masses,
whereas this is more difficult in the real world.
In the big bang the expansion of the universe is quite slow
compared to strong interaction scales, so equilibrium is
very nearly maintained, while in heavy ion collisions this
is much more doubtful.

We may expect there to be phase transitions in QCD, because
hadronic matter at zero temperature differs qualitatively from
what we expect at high temperature.  Asymptotic freedom implies
that at high temperature we shall have nearly free quarks
and gluons in a weakly interacting plasma.  On the other
hand at zero temperature the quarks and gluons are confined.
Also we know from a rich phenomenology of soft pion physics that
chiral $SU(2)\times SU(2)$ is slightly intrinsically but more
importantly spontaneously broken at zero temperature; the
spontaneous breaking will go away
at high temperature.

What can we say about the character of the transitions?  I
would like to organize the discussion by reference to the
following table.  The table presents a certain mixture of
established results, folk wisdom, and guesswork, as I shall
now explain.
\medskip
\tenpoint
\vbox{\tabskip=0pt \offinterlineskip
\def\tablerule{\noalign{\hrule}}
\halign to 6in{\strut#& \vrule#\tabskip=1em plus2em& #\hfil& \vrule#&
\hfil#\hfil& \vrule#& \hfil#\hfil&\vrule#&#\hfil&\vrule#\tabskip=0pt
\cr\tablerule
&&\multispan7\hfil&\cr
&&\multispan7\hfil CHARACTER OF PHASE TRANSITIONS \hfil&\cr
&&\multispan7\hfil&\cr\tablerule\tablerule
&&\omit\hidewidth {\bf Name}\hidewidth&&
\omit\hidewidth {\bf Behavior of}\hidewidth&&
\omit\hidewidth {\bf Phenomena}\hidewidth&&
\omit\hidewidth {\bf Examples}\hidewidth&\cr
&&&&{\bf Thermodynamics}&&&&&\cr\tablerule\tablerule
&&&&&&&&&\cr
&&1st order&&discontinuous&&latent heat&&F=0 deconfinement&\cr
&&&&&&&&&\cr
&& && &&supercooling&&f$\geq$3 chiral&\cr
&&&&&&&&&\cr\tablerule
&&&&&&&&&\cr
&&2nd order&&continuous, but&&long range&&f=2 chiral&\cr
&&&&not analytic&&correlations&&&\cr
&&&&&&&&&\cr
&&&&&&critical slowing&&&\cr
&&&&&&&&&\cr\tablerule
&&&&&&&&&\cr
&&quantum&&essential singularity&&mass gap&&deconfinement, F$\geq$1&\cr
&&&&at T=0&&&&&\cr
&&&&&&(P)$\sim e^{-M_{G}/T}$&&(ionization)&\cr
&&&&&&&&&\cr\tablerule
&&&&&&&&&\cr
&&none&&rapid but smooth&&quench fun&&F=2 chiral,&\cr
&&&&change&&&&$0<M_q\ll$ QCD&\cr
&&&&&&&&&\cr
&&&&&&&&F$\geq$1, very heavy quark&\cr
&&&&&&&&&\cr\tablerule \noalign{\bigskip}
&\multispan7{f$ \equiv \#$ of massless flavors}\hfil&\cr
&\multispan7{F=\# of flavors}\hfil&\cr
&\multispan7\hfil&\cr
}}
\twelvepoint

The first point is that a strict definition of
confinement is more elusive than one might expect intuitively.
It is instructive to compare, in this connection, the ionization
of ordinary gases as they are heated.  There is no question
that the plasma one obtains at high temperatures, with free
electrons and ions, behaves strikingly differently from the neutral
gas.  At Princeton, these two kinds of matter are studied on
different campuses.  However, there is no sharp phase
transition between them.  As the temperature rises, the amount
of ionization increases continuously,  One cannot identify a
specific temperature where the gas has suddenly become a plasma, and
the thermodynamic functions are perfectly smooth and continuous
(except at $T~=~0$, see below).  Likewise in the case of QCD,
one should not leap to the conclusion that deconfinement is
either easy to identify unambiguously,
or that it necessarily associated with a sharp phase transition.

In the pure glue version of QCD, with no quarks, it is
possible to give a useful strict definition of confinement,
using the Wilson-Polyakov loops.  In physical terms, one
considers the free energy in the presence of fixed sources of
color triplet charge, such as would be provided by very
heavy (static) test quarks.  If the correlation energy between
quark and antiquark
sources grows linearly with their separation for
large separation, we are in a confined phase.  Physically,
this arises because the quark source carries a quantum number,
triality, that is conserved modulo 3 and cannot be
screened by gluons which have zero triality.  If the
ground state is sensitive to the triality flux which connects
the sources, there will be a finite energy cost
per unit length
for the region of disturbed vacuum.  On the other hand
in more realistic versions of QCD, with quarks as dynamical
degrees of freedom, the correlation energy will remain finite
even as the sources are infinitely separated.  This is because
a single dynamical antiquark near the quark source can neutralize
its effect, and a single dynamic quark near the antiquark source
can neutralize its effect.  Thus the correlation energy cannot
be greater than a finite quantity, roughly speaking
twice the difference between the lightest meson containing a
heavy test quark and the bare mass of the heavy quark itself.

Thus for the pure glue theory there is a well-defined criterion
for confinement, namely the existence or not of a non-trivial
$Z_n$ symmetry.  The gauge invariant variables which transform
non-trivially under this symmetry are the loop integrals
$$
L(x) ~=~ {\rm Tr}~ P~\exp (i\int^\beta_0~ A_\tau (x)~d\tau )
$$
over imaginary time $\beta~=~T^{-1}$.  The expectation value
of such an integral vanishes in the confined phase, as discussed
above, while in the unconfined phase it does not vanish.
If we make the bold hypothesis that for considering the
phase transition it is only necessary to study the coarse-grained
variables, and that the only quantities that survive coarse-graining
are those connected with conserved quantities deriving from symmetries,
then
the deconfining transition
in the pure glue theory becomes related
to a much simpler dynamical model having the same
symmetry, namely a $Z_n$ gauge theory.
In turn, the
$Z_n$ gauge theory can be mapped by a duality
transformation to a simple global $Z_n$ theory:
the Ising model for $n~=~2$, and its straightforward
generalization, called the Potts model,
for $n~=~3$.  These models have been extensively studied both
analytically and numerically.  It is known that the
Ising model has a second order transition from its
magnetized to its unmagnetized state, while for the
Potts model the transition is first order.  Numerical
simulations of pure glue QCD show that
one has a second order deconfining transition for
$SU(2)$ and a first order transition for $SU(3)$, confirming
this predicted pattern.

I would like to comment
on the proper
logical structure of preceding important
argument [\sy ], whose true significance is
quite subtle.
Second-order transitions are characterized by
continuous but non-analytic behavior of important
thermodynamic quantities, including especially
the order parameter, near the transition temperature.
Referring to the definition
of the partition function, we see that non-analyticity
in the temperature $T$ can only arise (for $T\neq 0$)
when there is a subtlety in taking the infinite volume
limit.  In turn, such subtleties indicate that there
are important long-range correlations in the system, which
are responsible for the non-analytic behavior.  Thus in
studying the singularities near second-order phase transitions
we should be able to concentrate on models that describe only
the
long-wavelength modes -- scale-invariant theories, right
at the critical point.  The non-analytic part of the
thermodynamics near the critical temperature will
be largely insensitive to the detailed microscopic interactions
at short wavelengths.    This is the basic argument
for universality,
and although this argument
is not entirely rigorous it has been very
fruitful and successful in allowing one to
describe many second order transitions, following Wilson,
by studying the behavior of appropriate simple scale-invariant
models.  Models that are much simpler than the original
microscopic model of interest can nevertheless describe the nature of
its critical
singularities exactly, as long as the two models
have the same symmetries
and thus the same mode structure at long wavelengths.
So in examining the possibility that (any version of) QCD
has a
second order phase transition, we are invited to try to construct
a candidate scale-invariant theory with the same symmetry.
If we can, we will have constructed at least a logically
satisfactory model, becoming exact near the transition, for a
 second order QCD transition.
Thus in the preceding example, the existence of a second order
transition for the Ising model shows that this is a logical
possibility also for pure $SU(2)$ gauge theory.
Having a consistent model
does not in itself
prove that the transition of
interest in the microscopic model
actually is second order --
it remains logically
possible that its order parameter will simply change
discontinuously and never develop long correlation lengths.
On the other hand if we cannot find a suitable
scale-invariant model, then the transition cannot be second
order.  This is what happened for
the $SU(3)$ pure glue theory.
In this case too our footing is not entirely solid:
strictly speaking,
all that can be said is that the simplest candidate model, the
Potts model, does not work.  Clearly there is a strong
element
of guesswork in this general
approach to anticipating the nature
of phase transitions by searching for scale-invariant models.
Nevertheless it has been applied
successfully in many cases to condensed matter systems,
and now to the deconfinement transition for pure glue.

A first-order transition is robust to small perturbations, and
so one should expect that ($n=3$) QCD with only very massive quarks
would still have a first-order phase transition, the ghost
of the true confining transition for no quarks, although there
would no longer be a good order parameter for this transition.

\bigskip

There is a formal sense in which confinement represents
a zero-temperature phase transition, even in the presence
of dynamical quarks.
There is an essential
singularity at $T~=~0$, due to the existence of a mass
gap, which appears in the Boltzmann factor
$e^{-m_G/T}$.  This singularity does not appear in the free theory.
In fact the mass gap is non-perturbative in $g^2$, as
discussed above.

\bigskip

In the real world we have at least two flavors of quarks
that must be regarded as very light ($m_q << \Lambda_{\rm QCD}$),
and it is doubtful whether deconfinement shows itself even indirectly
in the
shadowy form of a first-order transition [\kw ].
Indeed
the lattice simulations
for two light quarks seem to indicate that while there is still a
phase transition in that case it has a drastically different
character from the pure glue transition.  The
transition for two massless quarks has a significantly
lower critical temperature
than in the quarkless case,
and much less latent heat -- possibly none.  It seems more
appropriate to identify the cause of the transition in the
two quark theory as being restoration of chiral symmetry,
as I shall now elaborate\foot{One should also consult
Leutwyler's talk in this volume,
where he presents an approach to the
phase transition based directly on chiral perturbation theory.
This approach is complementary to the one described below,
in that it is microscopically based and allows one to estimate
non-universal properties such as the critical temperature, but
cannot address the behavior very near the transition.}.  On the
other hand one expects a steep though smooth rise
in the free energy on passing through the Hagedorn temperature,
reflecting the possibility of producing highly excited resonances,
saturating at the gluon plasma value.  In fact in the
simulations the chiral transition
seems to occur at the startlingly
low temperature $T\approx 150{\rm MeV.}$, followed
by a steep but continuous rise in the free energy per unit volume
to a value five times as large at
only slightly higher temperatures.

We can repeat for chiral symmetry the exercise,
which supplied such excellent guidance in the pure
glue case, of surveying
the possibilities for second order transitions.
The simplest models
with the appropriate symmetries are the linear sigma model
for $SU(2)\times SU(2)$ chiral symmetry
and its matrix generalizations
for $SU(f)\times SU(f)$ for $f\geq 3$.

In the $f=2$ case
we are dealing with a model that has been studied
in great
depth for purposes condensed matter physics.
Indeed the fields
$\vec \phi \equiv (\sigma ,\vec \pi )$ in this model, subject
to an $SU(2)\times SU(2) \approx SO(4)$ symmetry, are just
the same fields one would introduce to describe the
magnitude and direction of an isotropic four-component
magnet.  It is known from extensive analytic and numerical
work that this model has a second order phase transition, and
its critical exponents have been calculated accurately.
Since there is a scale-invariant model with
the right symmetries,
there is the strong logical possibility of a second order
chiral transition for two-massless-flavor QCD.

On the other hand the corresponding models for $f\geq 3$ do
not give rise to any scale-invariant theories.  (In the
language of the renormalization group, they do not have
infrared stable fixed points.)  This was first
suggested by a crude calculation using the $\epsilon$-expansion,
and subsequently demonstrated by direct numerical simulation.
Subject to the same
{\it caveats\/}
mentioned above,
this analysis suggests that the chiral phase transition
in QCD is second order for two masses quarks but first order
for three or more.  This picture is quite consistent with existing
simulations.

Using the
concept of universality one can translate existing
results
for the magnet model
into a wealth of
precise predictions for the behavior of two-flavor
QCD near its transition, if it is second order.
These include detailed quantitative estimates of the
specific heat, the behavior of the vacuum expectation value
$<\bar q q>$ as a function of temperature and bare quark
mass, and many others.  For example one predicts that
the specific heat has a cusp of the form
$|T-T_c|^{-\alpha}$,
$\alpha \approx -.21$, with coefficients in the ratio
$A_+/A_- \approx -1.9$.  Another striking prediction
is that the $\sigma$ mass$^2$, defined to be the inverse
correlation function at zero momentum
in the scalar isoscalar channel,
vanishes at fixed temperature near $T_c$ like the
square root of the bare
quark mass as this mass is taken to zero.  This effect
is entirely due to the fluctuations; in mean field theory
the mass$^2$ remains finite.

One can also discuss the effect of
adding a third massive quark flavor,
which is quite interesting and
presumably brings the model quite close to real QCD.
If the strange quark were massless the transition
would be (according to
our expectations) first order, accompanied
by release of latent heat, while if
the strange quark were infinitely massive the transition
would be second order.  The simplest logical possibility is
that as the mass is increased the latent heat continuously
shrinks to zero, and the first-order transition joins continuously
to the second-order one.  The form of the thermodynamic
singularities near the {\it tricritical point\/} where the joining
occurs can also be calculated using renormalization group methods.
It is quite poetic that these singularities are governed by
a massless $\phi^6$ theory, which is asymptotically free in
three dimensions!  Thus at this point QCD exhibits scale
invariance up to calculable
logarithmic corrections both in the infrared and
in the ultraviolet.  Many specific predictions follow.  For
example, near the tricritical point the specific heat develops
a true discontinuity, as opposed to a cusp.

We can look forward to interesting numerical experiments
on QCD thermodynamics in the near future.
Many precise predictions
regarding the
suggested second-order and tricritical transitions
are waiting to be tested.

For possible cosmological applications, the situation seems less
exciting.
Various dramatic consequences that might have attended a strongly
first order QCD phase transition, including production of new forms
of matter or modification of standard nucleosynthesis scenarios,
almost certainly did not occur.

\REF\lasers{A. Anselm and M. Ryskin, {\it Physics Letters\/}
{\bf B226} 482 (1991); J.-P. Blaizot and A. Krzywicki,
{\it Soft Pion Emission in Heavy-Ion Collisions\/} Orsay
preprint LPTHE 92/11.  There is also important unpublished
work on the subject by J. Bjorken and by K. Kowalski and C.
Taylor.}

Likewise if heavy ion collisions can be described as an evolution
through near-equilibrium states, then no dramatic signature emerges.
{\it If\/} pions were truly massless, then near what
would then be a true
second-order transition
one would find extended
fluctuations with very long correlation
lengths and times.
In other words, there would be large
long-lived regions of misaligned vacuum, where $\vec \phi$ could
point in any possible given direction.
These regions would act as pion lasers [\lasers ],
emitting coherent pion radiation.
(Mathematically identical sources of Nambu-Goldstone
bosons radiation arise in cosmic texture models.)
Unfortunately in the real world pions are not massless, and
indeed their mass is probably not much less than $T_c$.  Thus
truly long correlation lengths do not develop, and the very
interesting possibilities just mentioned do not occur.

On the other hand it is far from clear that the assumption of
approximate equilibrium is reasonable in the context of heavy ion
collisions.  It may be no more unrealistic to consider the
opposite idealization, of a rapid {\it quench\/} from a very
high to zero temperature.   Under these conditions there can
be growth, at the speed of light,
of large domains of correlated field
which subsequently
relax {\it coherently\/} toward the true vacuum direction by
emitting coherent pion radiation.  The resulting clusters
of pions would be highly correlated in momentum and in charge,
and could conceivably give rise to a spectacular
phenomenology.

\chapter{Scaling Violations and the Running of the Coupling}

A certain misunderstanding of the connection between running of the
coupling and violations of scale invariance in QCD seems to be
prevalent, and I add this brief section to help dispel it.

The simple fact is that {\it all\/} violations of scale invariance
in QCD, aside from those due to explicit quark masses, indicate
running of the coupling.  Indeed with a fixed numerical value of the
-- dimensionless -- coupling (and zero quark masses) the Lagrangian
of QCD is explicitly and unavoidably scale invariant.  Such is the
character of the classical theory.  On the other hand, the quantum theory
is fully determined once one specifies the value of the coupling at some
reference momentum.  If the coupling were not to change
as the choice of
reference momentum was varied then there would be no violation of scale
invariance, just as in the classical theory.

These elementary but profound aspects of QCD are somewhat obscured in
the traditional presentations of the predictions of the theory.  It may
be instructive, therefore, to bring out more clearly the underlying
dependence
of these predictions
on the running coupling in the most important case of deep
inelastic scattering.    The primary result of
the standard analysis of the operator product expansion, applied to
deep inelastic scattering, is to express the moments of structure functions
as linear combinations of matrix elements of
the appropriate spin operators.
For simplicity let us consider a non-singlet
case and only the leading term at large $Q^2$, so that only one operator
of lowest twist contributes for each spin. It will be obvious that the
main point does not depend on these specializations.
The relevant equation is then
$$
\int^1_0 x^n F(x,Q^2) ~=~ \langle {\cal O}_n \rangle {\cal C}_n(Q^2)~.
\eqn\appsra
$$
Here the matrix element $\langle {\cal O}_n \rangle$ is independent of
$Q^2$.  Also, and most profoundly, ${\cal C}_n(Q^2)$ is a definite
function of $\bar g (Q^2)$, the effective coupling normalized at $Q^2$.
When it is so expressed,
it has no further $Q^2$ dependence.  This is
because the theory is completely specified once the coupling
at some definite mass scale is prescribed.  So once
the single mass scale at which the coupling is prescribed
is chosen
to be $Q^2$ itself -- or rather, to be precise, the
square root of $-Q^2$ -- the rest is pure numbers.  All other mass
scales, such as the masses of hadrons,
can be expressed as multiples of $Q^2$ times
definite numerical functions of $\bar g (Q^2)$.
Indeed, this phenomenon
is just the converse
of the dimensional transmutation we discussed previously:
here instead of
trading a coupling for a mass scale, we are trading a mass scale for
a coupling.

Thus for the ratio of moments at two values of $Q^2$ we have
$$
{\int^1_0 x^n F(x,Q_2^2) \over \int^1_0 x^n F(x,Q_1^2)}
{}~=~ {{\cal C}_n(\bar g (Q_2^2)^2 )\over {\cal C}_n(\bar g (Q_1^2)^2)}~.
\eqn\appsrb
$$
Clearly therefore {\it any\/} violation of scale invariance -- that is,
any variation of the left-hand side from unity -- reflects running of the
coupling quite directly.

To make contact with the traditional presentation one must solve the
differential equation for the evolution of ${\cal C}_n$ with the
coupling, to make the right-hand side of \appsrb\ more explicit.  For small
$\bar g$ the leading dependence is of the form
$$
{\partial\over \partial (\bar g (Q^2)^2 ) } {\cal C}_n(\bar g (Q^2)^2)
\approx c_n {\cal C}_n(\bar g (Q^2)^2)~,
\eqn\appsrc
$$
where $c_n$ is a calculable number.  This
form reflects that the scale
dependence on the coupling arises
as a quantum radiative correction.  Integrating, we have
$$
{\int^1_0 x^n F(x,Q_2^2) \over \int^1_0 x^n F(x,Q_1^2)}
{}~=~ {{\cal C}_n(\bar g (Q_2^2)^2) \over {\cal C}_n(\bar g (Q_1^2)^2)}
{}~\approx~ \bigl( {\bar g(Q_2^2)^2 \over \bar g(Q_1^2)^2 } \bigr)^{c_n}~.
\eqn\appsrd
$$
Then inserting the running of the coupling
$\bar g (Q^2)^2 \propto 1/\ln (Q^2/\Lambda^2)$ for large $Q^2$, we
reach the traditional form
$$
{\int^1_0 x^n F(x,Q_2^2) \over \int^1_0 x^n F(x,Q_1^2)}
{}~=~
\bigl( {\ln (Q_1^2 /\Lambda^2) \over \ln (Q_2^2 /\Lambda^2) }\bigr)^{c_n}~.
\eqn\appsre
$$
In this form the underlying dependence of the scaling violation on the
running of the coupling is hidden from view.
Nonetheless, as I hope this discussion
has made crystal clear, running of the coupling is the primary cause
of the scaling violation.

These comments are
in no way intended to deprecate the achievements of workers
who succeed
in extracting measures of the effective couplings
at different mass scales directly, for
example from studies of
$\tau$-decay or from jet physics at the Z resonance, and by
comparison can demonstrate directly and dramatically the running of the
coupling.  These are magnificent achievements, justly celebrated in the
following pages.  However I must insist on the important logical point
that in QCD
{\it all\/} scaling violations provide {\it prima facie\/} evidence
for running of the coupling.

\refout

\end